\documentclass[11pt]{article}
\usepackage{fullpage}
\usepackage{xspace}
\usepackage[ruled,vlined]{algorithm2e}
\usepackage{setspace}
\usepackage{mathtools}
\usepackage{amsmath}
\usepackage{amssymb}
 
\usepackage{amsfonts}
\usepackage{amsthm}
\usepackage[dvipsnames]{color}
\usepackage[stable]{footmisc} 

\usepackage{thmtools,thm-restate}
\usepackage{hyperref}
\usepackage{cleveref}
\usepackage{comment}
\usepackage{multicol}
\usepackage{multirow}
\usepackage[toc,page]{appendix}
\usepackage[affil-it]{authblk}
\usepackage{url}
\usepackage{authblk}

\usepackage{caption}
\usepackage{subcaption}

\usepackage{csquotes}
\usepackage{xspace}
\newcommand{\st}{~|~\xspace}

\renewcommand{\leq}{\leqslant}

\renewcommand{\geq}{\geqslant}

\newtheorem*{theorem*}{Theorem}

\newtheorem{corollary}{Corollary}

\begin{document}

\title{Simplicity in Eulerian Circuits:\\ Uniqueness and Safety}

\author[1]{Nidia Obscura Acosta}
\author[2]{Alexandru I. Tomescu}
\affil[1]{Department of Computer Science, Aalto University, {\tt nidia.obscuraacosta@aalto.fi}}
\affil[2]{Department of Computer Science, University of Helsinki, {\tt alexandru.tomescu@helsinki.fi}}

\date{}
\setcounter{Maxaffil}{0}
\renewcommand\Affilfont{\itshape\small}

\maketitle

\begin{abstract}
An Eulerian circuit in a directed graph is one of the most fundamental Graph Theory notions. Detecting if a graph $G$ has a unique Eulerian circuit can be done in polynomial time via the BEST theorem by de Bruijn, van Aardenne-Ehrenfest, Smith and Tutte, 1941--1951 (involving counting arborescences), or via a tailored characterization by Pevzner, 1989 (involving computing the intersection graph of simple cycles of~$G$), both of which thus rely on overly complex notions for the simpler uniqueness problem.

In this paper we give a new linear-time checkable characterization of directed graphs with a unique Eulerian circuit. This is based on a simple condition of when two edges must appear consecutively in all Eulerian circuits, in terms of cut nodes of the underlying undirected graph of~$G$. As a by-product, we can also compute in linear-time all maximal \emph{safe} walks appearing in all Eulerian circuits, for which Nagarajan and Pop proposed in 2009 a polynomial-time algorithm based on Pevzner characterization.
\end{abstract}

\section{Introduction}

\subsection{Background}

Finding an Eulerian circuit in a graph, namely a closed walk passing through every edge exactly once, is arguably the most famous problem in Graph Theory. Euler's theorem from 1741~\cite{euler1741solutio}, states:\footnote{Note that the if direction was proved only later, by Hierholzer in 1873~\cite{hierholzer1873moglichkeit}.}
\begin{displayquote}
\emph{A graph has an Eulerian circuit if and only if every node has the same number of in-neighbors and out-neighbors.}
\end{displayquote}

In this paper all graphs are directed, and we further assume without loss of generality that they are also weakly connected, in the sense that their underlying undirected graph is connected. For simplicity of presentation, we also assume that they have neither parallel edges nor self-loops, otherwise we can replace these by paths of length two.

The above characterization implies not only that we can check in linear-time if a graph is \emph{Eulerian} (i.e., it has an Eulerian circuit), but we can also find an Eulerian circuit in linear time: when arriving with an in-coming edge $(u,v)$ to a node $v$, there is at least one unused out-going edge $(v,w)$, and any such out-going edge can be used to continue constructing the circuit, by a suitable representation of circuits using doubly-linked lists and keeping track of nodes with unused edges (see e.g.,~Hierholzer's algorithm~\cite{hierholzer1873moglichkeit,fleischner1990eulerian}).

This choice among out-going edges also means that the graph may admit multiple Eulerian circuits. Indeed, another classical result (for directed graphs) is that the number of Eulerian circuits can be computed in polynomial time, with the BEST theorem by de Bruijn, van Aardenne-Ehrenfest, Smith and Tutte~\cite{BESTtheorem,tutte1941unicursal}, from 1941--1951. This theorem states that the number $\epsilon(G)$ of Eulerian circuits in the directed Eulerian graph $G = (V,E)$ equals
\[\epsilon(G) = t(G) \prod_{v \in V}(d(v)-1)!,\]
where $d(v)$ is called the \emph{degree} of $v$ and equals the out-degree (equivalently, in-degree) of $v$, and $t(G)$ denotes the number of arborescences of $G$ (spanning directed trees of the directed graph $G$) rooted at any fixed arbitrary node of $G$ and directed towards that node. The quantity $t(G)$ can be computed via Kirchoffs matrix-tree theorem~\cite{1847AnP...148..497K} by computing matrix determinants.

In some applications, where some unknown object to be reconstructed can be modeled as an Eulerian circuit, one is also interested in checking whether the graph has a \emph{unique} Eulerian circuit, in order to be certain that the reconstructed object is indeed the correct one. For example, in Bioinformatics, Eulerian circuits are a theoretical model of genome assembly, see e.g.,~the introductory textbook by Waterman~\cite{waterman2018introduction}. While the BEST theorem can be applied to check whether $\epsilon(G) = 1$, Pevzner proved in 1989~\cite{pevzner1989tuple} a direct characterization of graphs with a unique Eulerian circuit. We cite below this characterization as stated by Waterman in the textbook~\cite[Theorem~7.5]{waterman2018introduction}:

\begin{displayquote}
\emph{Graph $G$ has a unique Eulerian circuit if and only if the intersection graph $G_I$ of simple cycles from $G$ is a tree.}
\end{displayquote}

The undirected graph $G_I$ is obtained by decomposing $G$ into simple cycles $c_i = v_1^{i}, v_2^{i}, \dots, v_k^{i}$ (i.e., cycles with all nodes distinct except for $v_1^{i} = v_k^{i}$). Nodes in $G$ might belong to several such cycles, but each edge can be used in at most one cycle $c_i$. We add a node $C_i$ to $G_I$ for each such cycle $c_i$ obtained from $G$. We then add an undirected edge between two nodes $C_i$ and $C_j$ in $G_I$ for each node contained in both $c_i$ and $c_j$ in $G$. See \Cref{fig-nagarajan} for an example and~\cite[Section 7.2]{waterman2018introduction} for more details. Note that if $G_I$ is a tree, then such decomposition of $G$ is also unique.

Even though this does not rely on the BEST theorem, ultimately it is not very different, since this tree will then correspond to the unique arborescence of $G$ (i.e., giving $t(G) = 1$). Indeed, assuming $G_I$ is a tree, we can take an arbitrary node $v$ of $G$, and observe that there is a unique path from any other node $u$ to $v$, implying a unique arborescence rooted at $v$ and directed towards $v$. This holds since any node in the same cycle $c_i$ as $v$ has a unique path to it in $c_i$ and for any node $u$ not in $c_i$ there is a unique path from any node in its cycle $c_j$ to any node in cycle $c_i$, as $G_I$ is a tree.

Since in practice the input graph may have more than one Eulerian circuit, we may settle for less when reconstructing the unknown object. Namely, we may report instead those walks that are subwalks of \emph{all} Eulerian circuits. By definition these are also part of the unknown Eulerian circuit, and thus correct for the application at hand. The idea of finding such partial solutions common to all solutions to a problem has appeared concurrently in several fields, such as Bioinformatics (see e.g.~\cite{DBLP:journals/jcb/IduryW95}, and almost all state-of-the-art genome assembly programs), and combinatorial optimization (see e.g.~\cite{persistency} for edges common to all maximum bipartite matchings). Recently, such partial solutions have been called \emph{safe}~\cite{DBLP:conf/recomb/TomescuM16}, and a series of papers proposed algorithms finding all safe partial solutions for other problems. For example, \cite{DBLP:conf/recomb/TomescuM16,DBLP:journals/talg/CairoMART19,DBLP:conf/icalp/CairoRTZ21,DBLP:conf/stacs/Cairo0RSTZ23} gave characterizations and optimal algorithms for the walks appearing in all edge-covering circuits of a strongly connected graph (that cover each edge \emph{at least} once, not \emph{exactly} once). Note that safe walks for edge-covering circuits are also safe for Eulerian circuits (by definition), but they are not all the safe walks. Recently,~\cite{DBLP:conf/recomb/KhanKCWT22} characterized the paths appearing in all flow decompositions of a flow in a directed acyclic graph. Note that an Eulerian circuit in a graph also induces a flow if we assign flow value 1 to every edge, however the result of~\cite{DBLP:conf/recomb/KhanKCWT22} is restricted to acyclic graphs.

In 2009, Nagarajan and Pop~\cite{DBLP:journals/jcb/NagarajanP09} proposed the first algorithm for finding safe walks for Eulerian circuits, in a brief note on page 901:

\begin{displayquote}
\textit{In the case of Eulerian tours [in our terminology: circuits], reconstructing sub-tours [in our terminology: sub-walks] that are part of every Eulerian tour is feasible in polynomial time (see Theorem 7.5 in Waterman, 1995) based on finding acyclic subgraphs in the cycle-graph decomposition of the original graph.}
\end{displayquote}

In the above, the \emph{cycle-graph decomposition} is the graph $G_I$ from \cite[Theorem 7.5]{waterman2018introduction}. Even though the intuition of the authors is the correct one, we argue that this note is incomplete. First, since $G_I$ may not be a tree, it may not be unique, a fact which is overlooked in the above note. Second, not any acyclic subgraph corresponds to a walk appearing in all Eulerian circuits, which we illustrate in \Cref{fig-nagarajan}. We believe that the authors meant those acyclic subgraphs of $G_I$ with the additional property that none of their edges are contained in a bi-connected component in $G_I$. Both of these issues suggest that this may not be the ``right'' characterization of such walks.  Moreover, even though this characterization can be fixed, it is still based on the intersection graph $G_I$ of simple cycles of $G$, thus leading to an algorithm more complex than necessary. 

%

\setcounter{figure}{-1}
\begin{figure}
\begin{subfigure}[b]{0.45 \linewidth}
\centering
\includegraphics[scale=0.3]{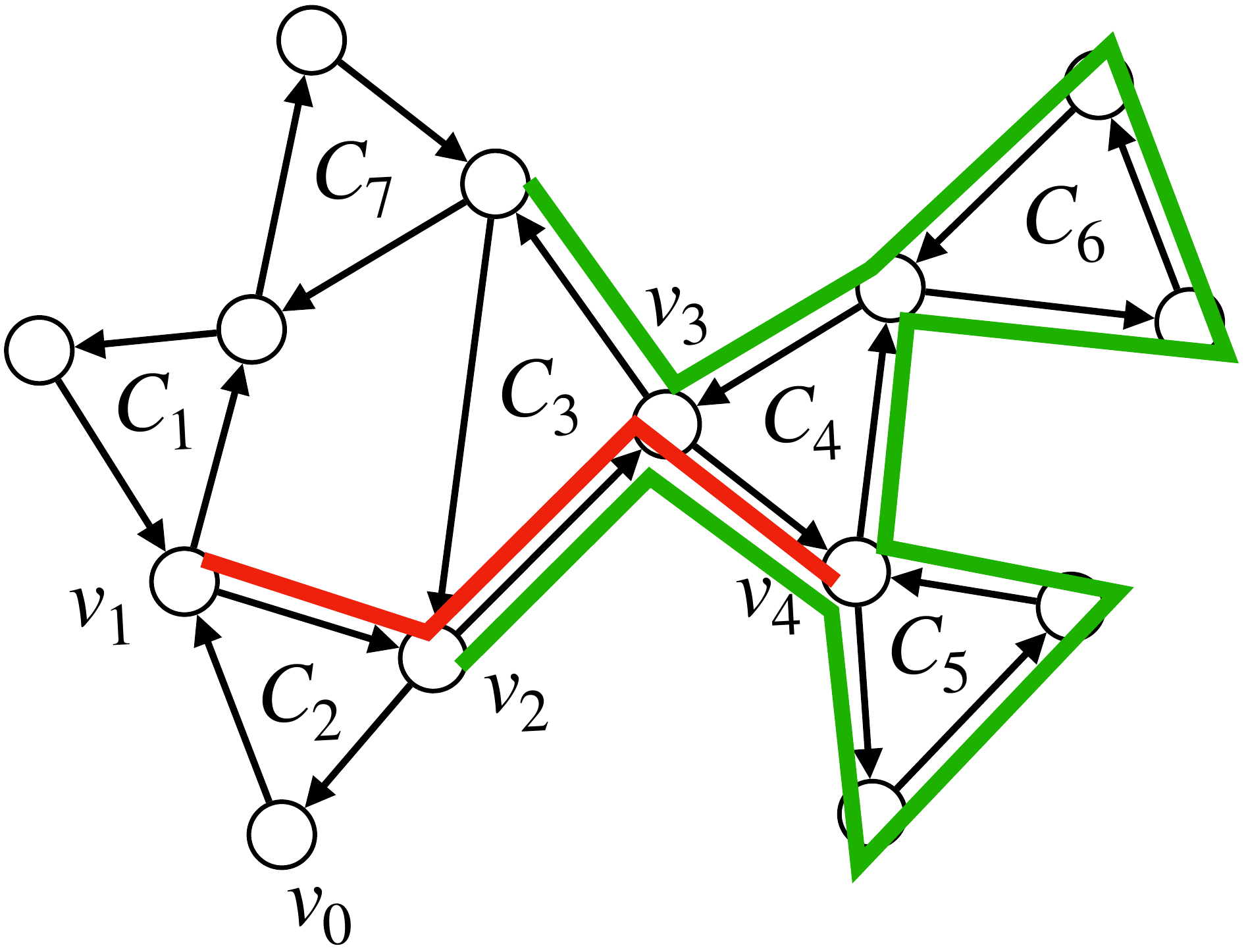}
\caption{An Eulerian graph $G$ and its simple cycles $C_1,\dots,C_7$\label{fig-nagarajan-1}}
\end{subfigure}
\hfill
\begin{subfigure}[b]{.45\linewidth}
\centering
\includegraphics[scale=0.3]{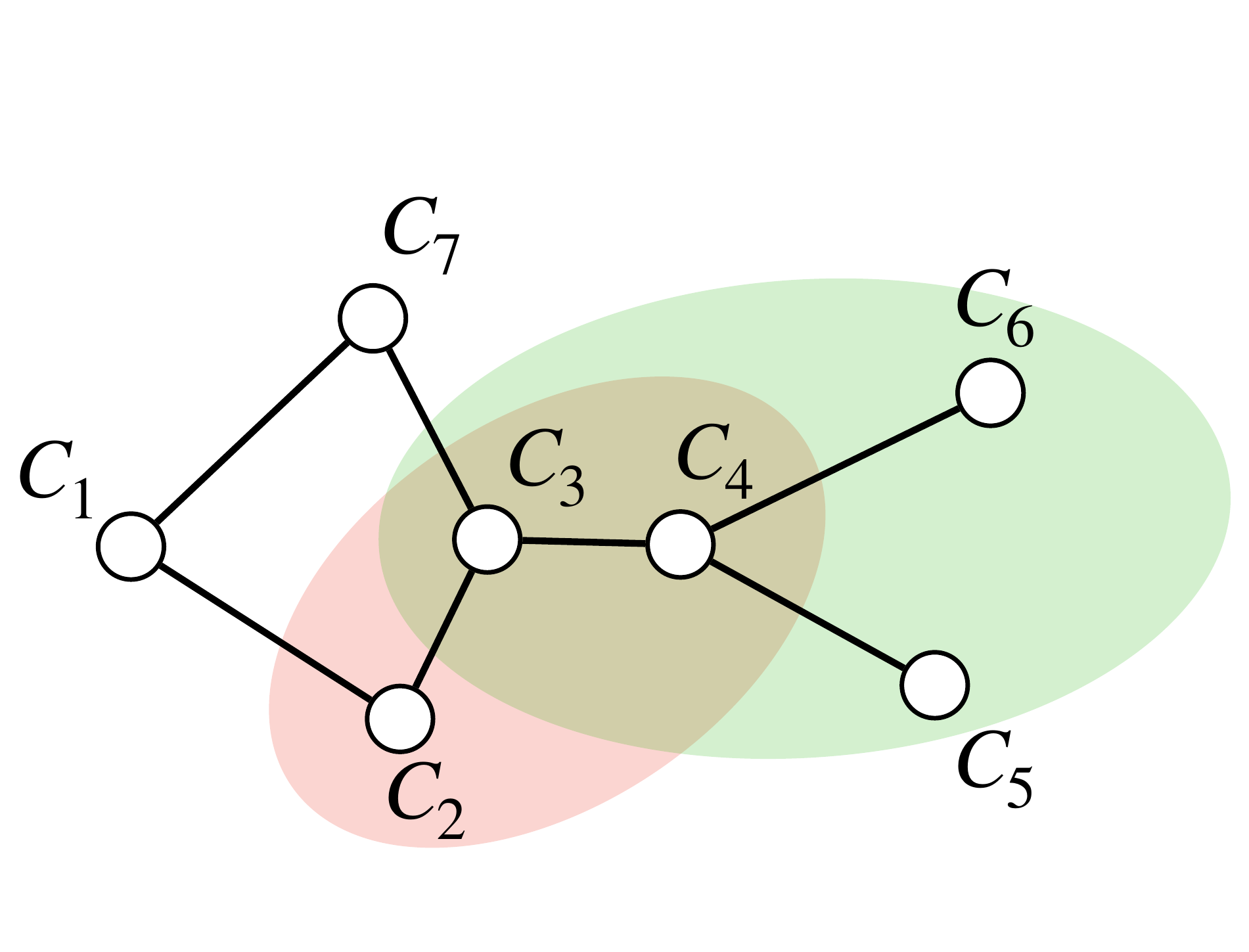}
\caption{An intersection graph $G_I$ of the simple cycles of $G$}
\end{subfigure}
\captionof{figure}{The acyclic subgraph of $G_I$ induced by $\{C_2, C_3, C_4\}$ (in red) has the property that no walk passing through cycles $C_2, C_3$ and $C_4$ in $G$ (i.e., using edges from each of them) is contained in all Eulerian circuits in $G$: the only such walk $(v_1,v_2,v_3,v_4)$, marked in red, is not contained in the Eulerian circuit which starts with $v_1, v_2$, then goes instead to $v_0$ and $v_1$ and continues to visit the rest of the graph, and finally back to $v_1$. However the acyclic subgraph of $G_I$ induced by $\{C_3, C_4, C_5,C_6\}$ indeed gives a walk in $G$ (marked in green) contained in all Eulerian circuits of $G$.\label{fig-nagarajan}}
\end{figure}

\subsection{Our Contribution}

In this paper we simplify both Pevzner's characterization of graphs with a unique Eulerian circuit~\cite{pevzner1989tuple} (also presented in Waterman's textbook~\cite{waterman2018introduction}), and the incomplete characterization of Nagarajan and Pop of safe walks for Eulerian circuits from~\cite{DBLP:journals/jcb/NagarajanP09}. Our idea is to characterize when an edge $(u,v)$ is followed by an edge $(v,w)$ in \emph{all} Eulerian circuits. Clearly, if $d(v) = 1$, this is always the case, and if $d(v) \geq 3$, it can be easily proved that this is never the case, see \Cref{fig:proof}(a) (which is also a simple idea behind the BEST theorem). The interesting case is thus when $d(v) = 2$. Let $U(G)$ denote the underlying undirected graph of $G$, namely the graph obtain from $G$ by removing the orientation of the edges; note that $U(G)$ is connected since we assume that $G$ is weakly connected. For any node $v$, denote by $U(G) \setminus v$ the graph obtained from $U(G)$ by removing $v$ together with all incident edges of $v$. We say that $v$ is a \emph{cut node} of $U(G)$ if $U(G) \setminus v$ is not connected.
We prove the following result:

\begin{theorem*}
\label{thm:main}
Let $G = (V,E)$ be an Eulerian graph, and let $(u,v),(v,w) \in E$. The walk $(u,v,w)$ is safe for Eulerian circuits (i.e.~it is a subwalk of all Eulerian circuits) if and only if it is a subwalk of an Eulerian circuit of $G$, and $d(v) = 1$ or $v$ has $d(v) = 2$ and is a cut node of $U(G)$.
\end{theorem*}

For example, in \Cref{fig-nagarajan}, $d(v_3) = 2$ and $v_3$ is a cut node of $U(G)$. Consequently, given $G = (V,E)$, we can define the set \[A(G) = \Big\{v \in V \st \text{$d(v) = 1$}~\vee~\big(\text{$d(v) = 2$ $\wedge$ $v$ is a cut node of $U(G)$}\big) \Big\},\]
and obtain a very simple characterization of graphs with a unique Eulerian circuit, not based on arborescences (as in the BEST theorem), nor on the intersection graph of simple cycles (as in Pevzner's characterization):

\begin{corollary}
Let $G = (V,E)$ be an Eulerian graph. We have that $G$ admits a unique Eulerian circuit if and only if $A(G) = V$. Moreover, we can detect if this is the case in time $O(|E|)$.
\label{cor:uniqueness}
\end{corollary}

Indeed, if $A(G) = V$, once we enter a node $v$ with some edge, we are forced to leave $v$ with a prescribed other edge, and since an Eulerian circuit visits every edge exactly once, the Eulerian circuit is uniquely determined.
The complexity bound follows from the fact that we can compute cut nodes in linear-time~\cite{hopcroft1973algorithm}, and this is the only non-degree based condition to be checked.

While at the beginning of the paper we mentioned that parallel edges and self-loops can be assumed to be absent (because they can be replaced by paths of length two), their presence has a very simple effect on \Cref{cor:uniqueness}. Parallel edges in an Eulerian graph clearly imply the presence of at least two Eulerian circuits, and self loops are allowed only for nodes having exactly one other outgoing edge.

In a similar way we can also obtain a simple and \emph{linear-time} algorithm reporting all maximal safe walks (i.e., not subwalks of other safe walks). 

\begin{corollary}
\label{cor:safety}
Let $G = (V,E)$ be an Eulerian graph. We can compute all maximal safe walks for the Eulerian circuits of $G$ (i.e., maximal subwalks of all Eulerian circuits of $G$) in time $O(|E|)$.
\end{corollary}

\begin{figure}[t]
\centering
\begin{subfigure}[t]{0.45\linewidth}
\centering
\includegraphics[scale=0.3]{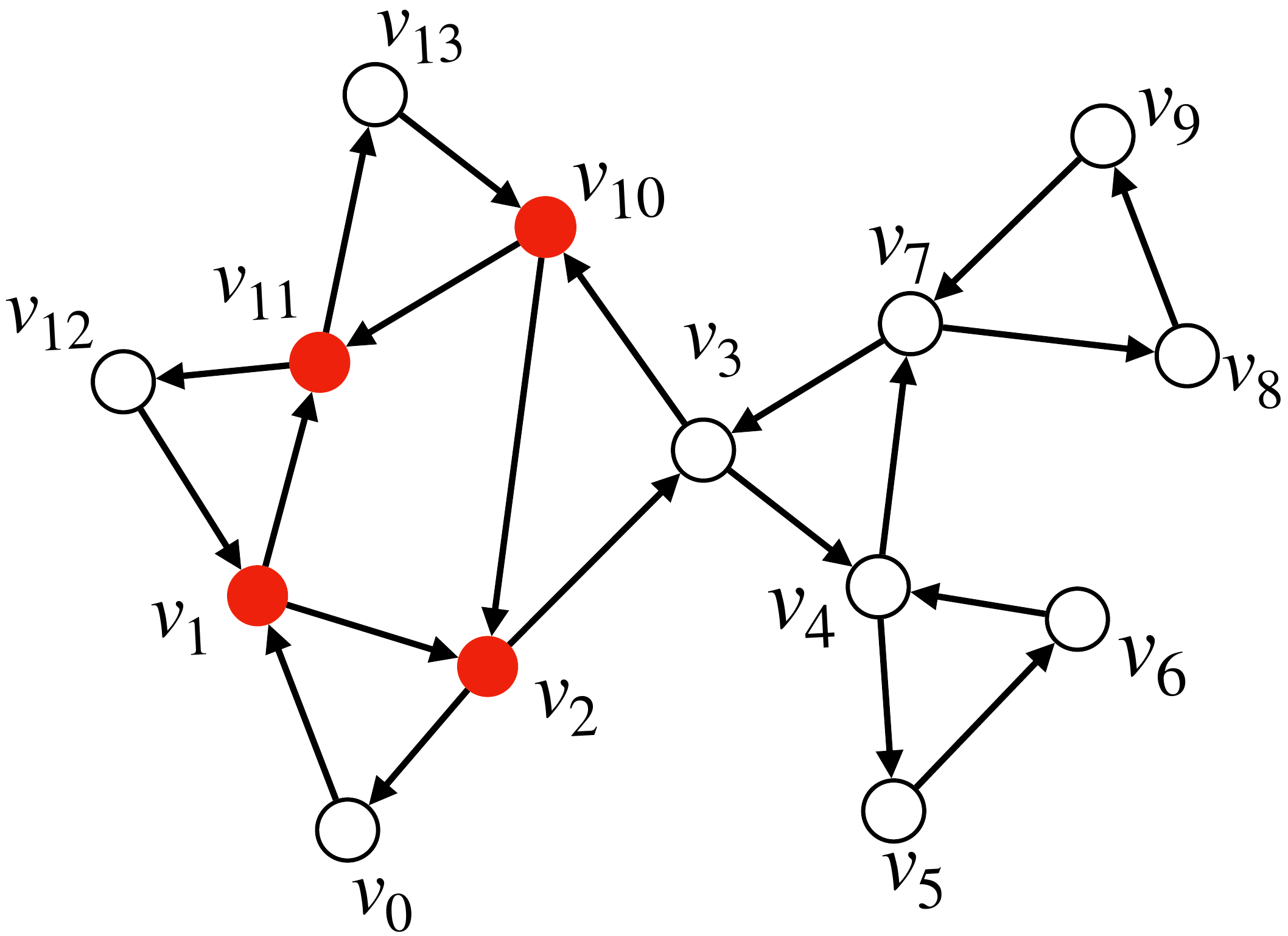}
\caption{The Eulerian graph $G$ from \Cref{fig-nagarajan}(a) with all nodes labeled}
\end{subfigure}
\hfill
\begin{subfigure}[t]{0.45\linewidth}
\centering
\includegraphics[scale=0.37]{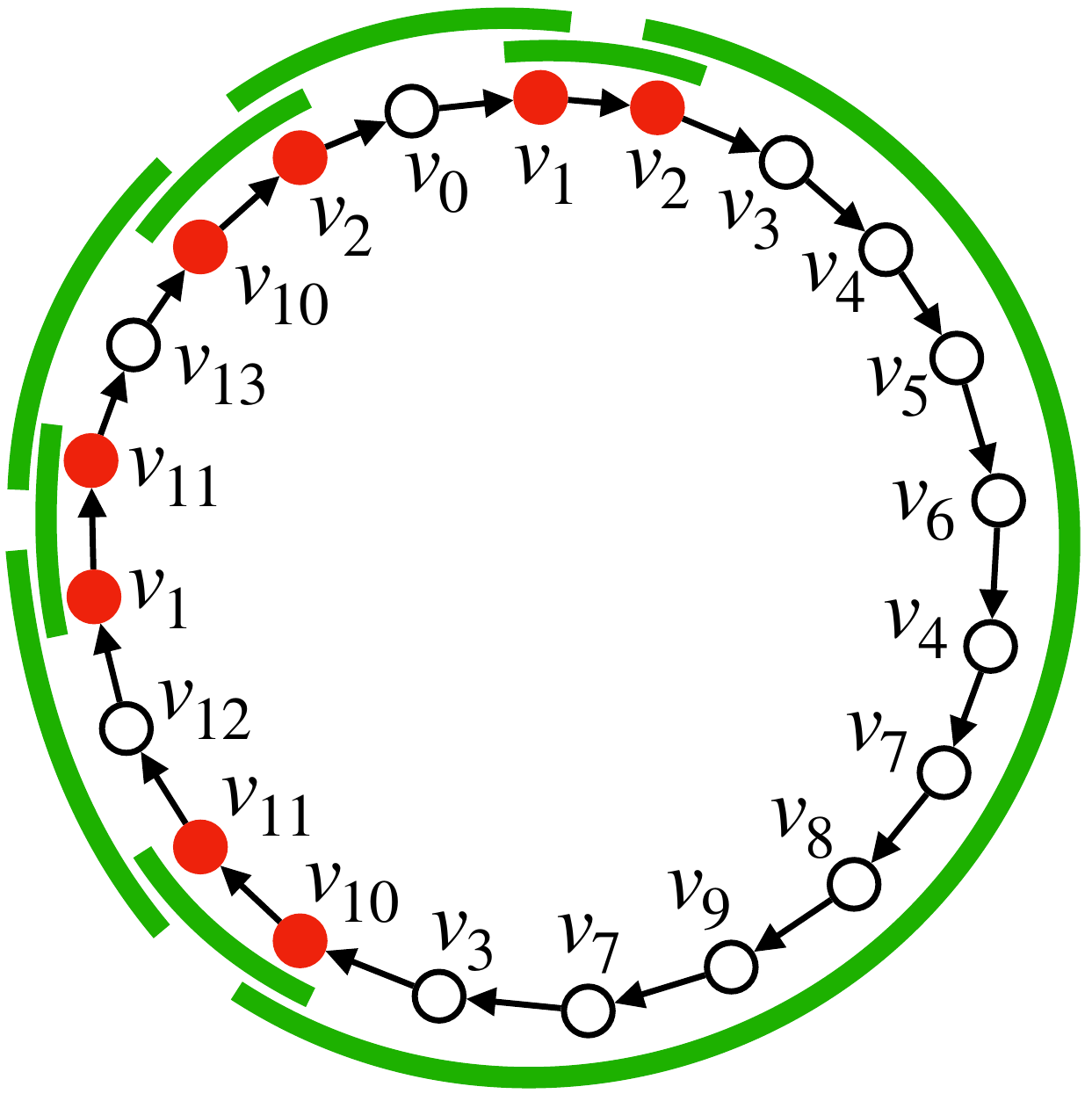}
\caption{An Eulerian circuit of $G$ and the maximal safe walks of $G$}
\end{subfigure}
\caption{Illustration of \Cref{cor:safety}. \Cref{fig:maximal}(a): the graph $G$ from \Cref{fig-nagarajan}(a) with all nodes labeled. It has $V(G) \setminus A(G) = \{v_1, v_2, v_{10}, v_{11}\}$ (shown in red). These four nodes define the cutting points for maximal safe walks in any Eulerian circuit of $G$. \Cref{fig:maximal}(b): one Eulerian circuit of $G$, the occurrences of $\{v_1, v_2, v_{10}, v_{11}\}$ in it (in red), and the maximal safe walks of $G$ (in green).} 
\label{fig:maximal}
\end{figure}

The above corollary is obtained by simply computing an Eulerian circuit $C$, in time $O(|E|)$, and cutting $C$ at any node $v$ not in $A(G)$ by keeping a copy of $v$ as an endpoint in each resulting segment (see \Cref{fig:maximal} for an example).
As argued above, each such maximal segment with internal nodes in $A(G)$ is a maximal safe walk because Eulerian circuits visit each edge exactly once.
Another immediate consequence is that maximal safe walks do not overlap on edges, and since every edge is safe, have total length exactly $|E|$, which is harder to observe without our theorem. This fact also makes the algorithm run in both input- and output-linear time.

\subsection{Notation}

Let $G = (V,E)$ be a directed graph. We denote $V$ by $V(G)$ and $E$ by $E(G)$.
Given $v\in V(G)$, we denote by $d(v)$ the number of out-neighbors of $v$ in $G$ (if $G$ is Eulerian, $d(v)$ also equals the number of in-neighbors of $v$).
We say that $W = (v_0,\dots,v_{k})$ is a \emph{walk} in $G$ if there is an edge from $v_i$ to $v_{i+1}$, for all $i \in \{0,\dots,k-1\}$ (the nodes of $W$ do not have to be distinct). We say that $W$ is \emph{non-empty} if $k > 0$. We say that $W$ is a \emph{circuit} if $v_0 = v_{k}$. Let $Q = (u_0,\dots,u_{\ell})$ be a walk with $\ell \leq k-1$. We say that $Q$ \emph{appears} in $W$ if $Q$ is a subwalk of $W$, in terms of edges (i.e. $Q$, viewed as a string of edges, is a substring of $W$, viewed as a string of edges). Similarly, if $W$ is a circuit, we say that $Q$ \emph{appears} in $W$ if there is an index $i \in \{0,\dots,k-1\}$ such that 
$(u_0, u_1) = (v_i,v_{(i+1) \bmod k})$, $(u_{1},u_2) = (v_{(i+1) \bmod k},v_{(i+2) \bmod k})$, \dots, $(u_{\ell-1},u_{\ell}) = (v_{(i+\ell-1) \bmod k},v_{(i+\ell) \bmod k})$.
Since we assume $G$ has neither parallel edges nor self-loops (since we can replace them by paths of length two), we can state the above two notions of ``appearance'' more simply just in terms of nodes.\footnote{We chose to define appearance in terms of edges (and not of nodes) in order to comply with the previous work on the safety question for Eulerian circuits~\cite{DBLP:journals/jcb/NagarajanP09}. Note however that safe walks are different depending on this choice. For example, assume $e_1$ and $e_2$ are two parallel edges from a node $a$ to a node $b$. Then the walk $(a,b)$ appears in all walks from $a$ to $b$, in terms of nodes, but neither $e_1$ nor $e_2$ appears in all walks from $a$ to $b$.}

For a directed or undirected graph $H$, and a node $v \in V(H)$, we denote by $H \setminus v$ the graph obtained from $H$ by removing $v$ together with all incident edges of $v$. If $H$ is undirected and connected, we say that $v$ is a \emph{cut node} of $H$ if $H \setminus v$ is not connected. For a set $K \subseteq V(H)$, we denote by $H[K]$ the subgraph of $H$ induced by $K$, that is, the subgraph of $H$ obtained by deleting all nodes not in $K$.

\section{Proof of the Theorem}
\label{sec:proofs}

\begin{figure}[t]
\begin{subfigure}[b]{.30\linewidth}
\centering
\includegraphics[scale=0.3]{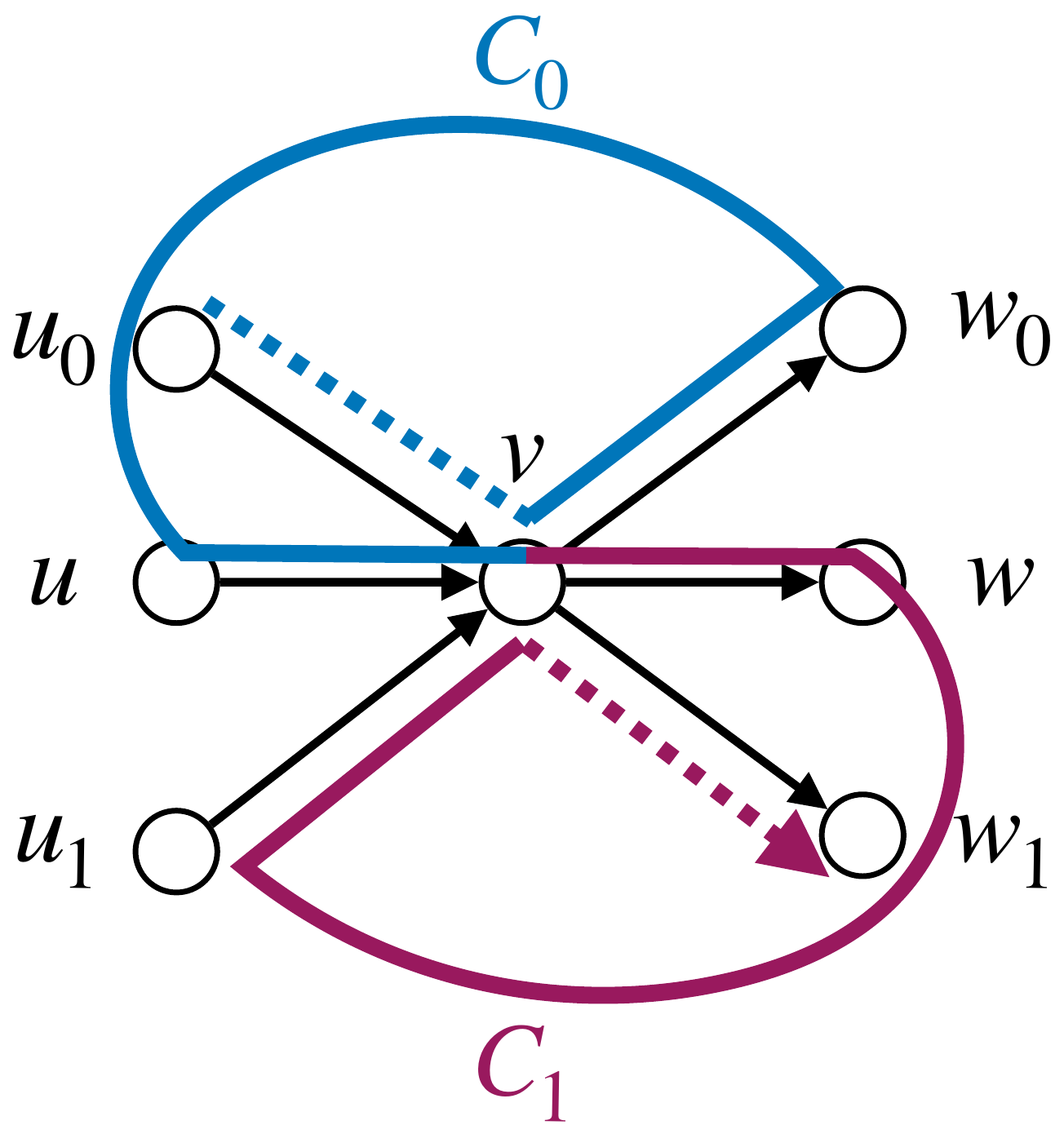}
\caption{Case 1, before swap\label{fig:case-1a}}
\end{subfigure}
\begin{subfigure}[b]{.30\linewidth}
\centering
\includegraphics[scale=0.3]{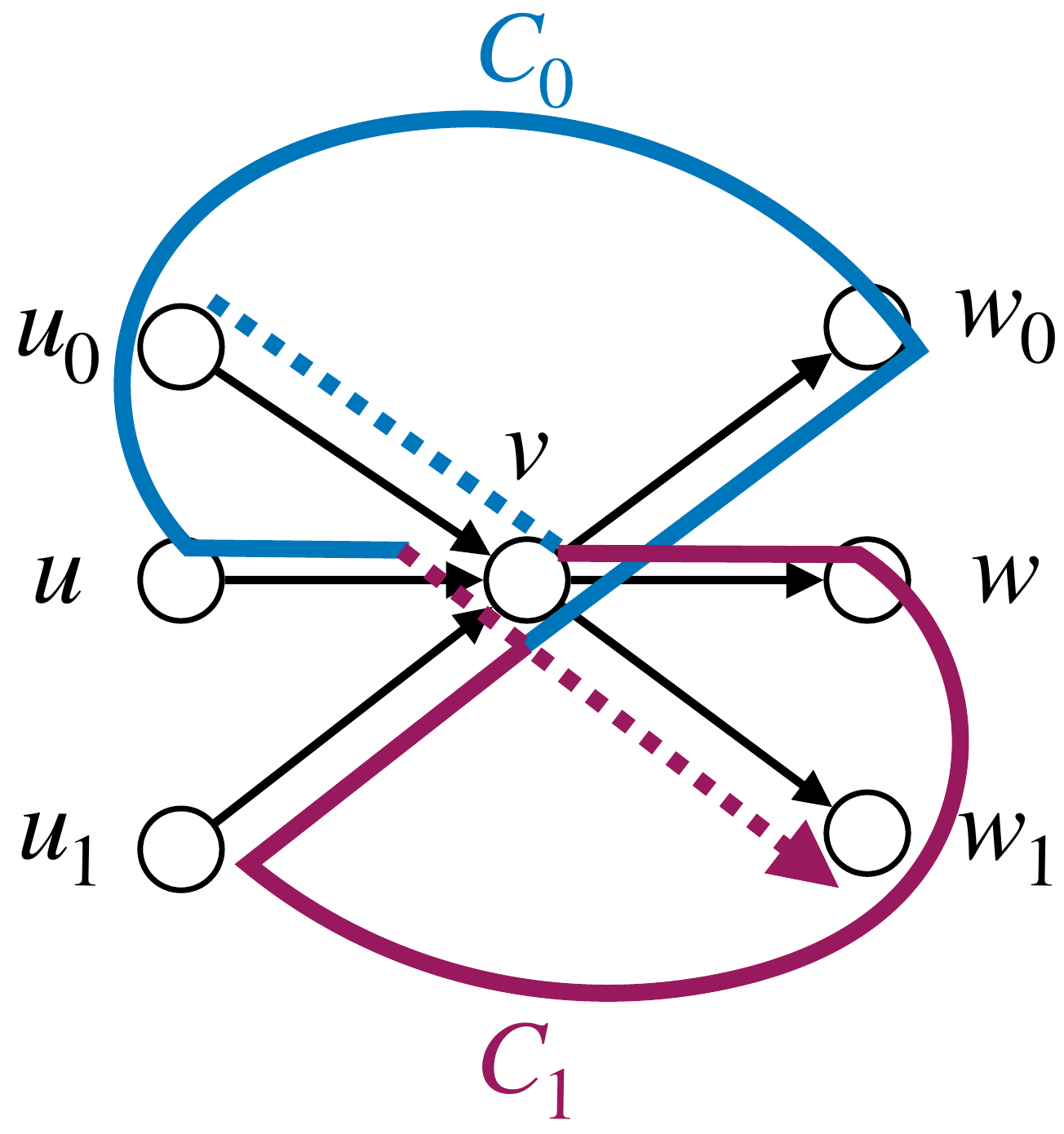}
\caption{Case 1, after swap\label{fig:case-1b}}
\end{subfigure}
\begin{subfigure}[b]{.30\linewidth}
\centering
\includegraphics[scale=0.3]{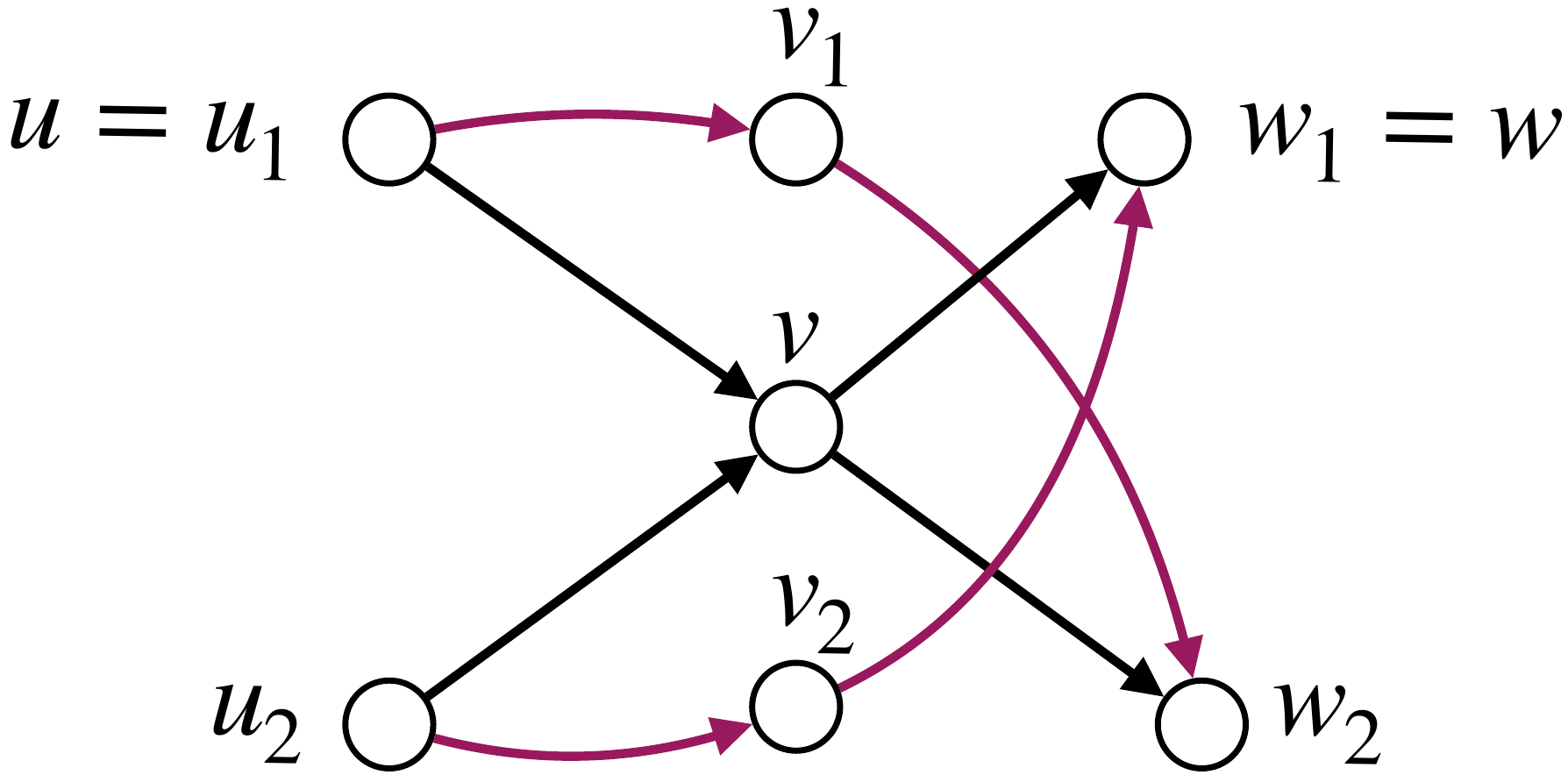}
\caption{Case 2\label{fig:case-2}}
\end{subfigure}

\caption{Illustration of the two cases in the proof of the theorem. \Cref{fig:case-1a}: Node $v$ has degree~3, and the Eulerian circuit $C$ visits $v$ with edge $(u_0,v)$ (dashed blue), then follows the walk $C_0$ (solid blue), then the walk $C_1$ (solid violet) and then the edge $(v,w_1)$ (dashed violet). \Cref{fig:case-1b}: By swapping $C_0$ and $C_1$ one obtains an Eulerian circuit $C^\ast$ in which $(u,v,w)$ does not appear. \Cref{fig:case-2}: Graph $G'$ constructed as in Case 2. Node $v$ is not a cut node and its degree is 2. The original edges are in black, and the newly added length-two paths are in violet; walk $(u,v,w)$ is not safe. \label{fig:proof}
}
\end{figure}

In order to be self-contained and to show that our arguments do not rely on complex results, we prove our theorem without using the BEST theorem. The only arguments are based on simply swapping parts of an Eulerian circuit (\Cref{fig:case-1a,fig:case-1b}), and a local construction when removing a non-cut node (\Cref{fig:case-2}).

\begin{proof}
We being by assuming that $(u,v,w)$ is a subwalk of all Eulerian circuits in an Eulerian graph $G$. Let $C$ be an Eulerian circuit of $G$, containing thus $(u,v,w)$. We will show that $d(v) = 1$ or $v$ has $d(v) = 2$ and is a cut node of $U(G)$ (equivalently, that $v \in A(G)$). Assume for a contradiction that $v \notin A(G)$; we have two cases:

\emph{Case 1.} Node $v$ has $d(v) \geq 3$. From $C$, we construct another Eulerian circuit $C^{\ast}$ which does not contain $(u,v,w)$, as follows. Denote by $v_0$, $v$, and $v_1$ three consecutive occurrences of $v$ in $C$, which exist because $d(v) \geq 3$. Let $u_0,w_0$, and $u_1,w_1$ be the nodes before, and after $v_0$ and $v_1$, respectively, on $C$. Let $C_0 = (v_0,w_0,\dots,u,v)$ and $C_1 = (v,w,\dots,u_1,v_1)$ be the subwalks of $C$ between $v_0$ and $v$, and between $v$ and $v_1$, respectively. We can obtain the circuit $C^\ast$ by swapping the two occurrences of $C_0$ and $C_1$ in $C$ (see \Cref{fig:case-1a,fig:case-1b}). Since $C^\ast$ is still a circuit and has the same set of edges as $C$, it is still Eulerian. Moreover, notice that in $C^\ast$ the walk $(u,v,w)$ does not appear, because edge $(u,v)$ is followed by edge $(v,w_1)$. This contradicts the fact that $(u,v,w)$ appears in all Eulerian circuits of $G$.

\emph{Case 2.} The degree of $v$ is $d(v) = 2$ and $v$ is not a cut node of $U(G)$. Let $u_1 = u$ and $u_2$ be the in-neighbors of $v$ and let $w_1=w$ and $w_2$ be the out-neighbors of~$v$. Consider the subgraph $G[V \setminus \{v\}]$ and note that, since $v$ is not a cut node of $U(G)$, we have that $U(G[V \setminus \{v\}])$ is connected. Let $G'$ be the graph obtained from $G[V \setminus \{v\}]$ by adding new length-two paths $(u_1,v_1,w_2),(u_2,v_2,w_1)$, where $v_1$ and $v_2$ are new nodes (see \Cref{fig:case-2}). Note that in~$G'$ there are neither parallel edges nor self-loops, every node has an equal number of in-neighbors and out-neighbors, and since $U(G[V \setminus \{v\}])$ is connected, $G'$ contains an Eulerian circuit $C'$. Note that $C'$ can be transformed into an Eulerian circuit $C^\ast$ of $G$, as follows: whenever $C'$ passes through a new path $(u_i,v_i,w_{3-i})$, we make $C^\ast$ pass through $(u_i,v,w_{3-i})$. The Eulerian circuit $C^\ast$ thus constructed does not pass through $(u=u_1,v,w_1=w)$. Thus, $(u,v,w)$ does not appear in all Eulerian circuits of $G$, a contradiction.

In order to prove the converse direction, assume that $v \in A(G)$ and $(u,v,w)$ is a subwalk of an Eulerian circuit in an Eulerian graph $G$. We will show that $(u,v,w)$ is safe, i.e. that it is a subwalk of all Eulerian circuits in $G$. If $d(v) = 1$, the claim obviously holds. Assume that $d(v) = 2$ and it is a cut node of $U(G)$. Thus, $U(G) \setminus v$ has exactly two connected components, $K_1$ and $K_2$, and $v$ has exactly one in-neighbor and one out-neighbor in each of $K_1$ and $K_2$. The only way for an Eulerian circuit $C$ to visit edges of $K_1$ or $K_2$ is using node $v$.
Therefore, immediately after reaching $v$ from its in-neighbor in $K_t$ (for $t \in \{1,2\}$), any Eulerian circuit $C$ must go to the other component $K_{3-t}$, because it is the only point when it can traverse it. This means that $u$ and $w$ do not belong to the same connected component $K_t$ (because $(u,v,w)$ appears in an Eulerian circuit), and any Eulerian circuit visits the edge $(v,w)$ immediately after visiting the edge $(u,v)$.
\end{proof}

\section{Conclusion}

In this paper we considered the problem of detecting when a graph has a unique Eulerian circuit, and the related problem of finding safe walks for Eulerian circuits. We simplified both existing characterizations, which are based on the intersection graph of simple cycles of~$G$. Inspired by the safety paradigm, we characterized when two edges must be consecutive in all Eulerian circuits, whose single non-trivial condition involves cut nodes of the underlying undirected graph.

This leads to very simple and linear-time detection of graphs with a unique Eulerian circuit, and linear-time computation of maximal safe walks for Eulerian circuits. The latter result also fits a line of research on walks safe for related objects, such as the walks appearing in all edge-covering circuits, or in all flow decompositions of a directed cyclic graph. The resulting linear-time complexity bounds are optimal and match the one for computing an Eulerian circuit.

\section{Acknowledgement}

We are very grateful to the anonymous reviewers who helped improved the presentation of this paper. This work was partially funded by the European Research Council (ERC) under the European Union's Horizon 2020 research and innovation programme (grant agreement No.~851093, SAFEBIO) and partially by the Academy of Finland (grants No.~322595, 328877, 314284 and 335715).

\bibliography{bibliography}

\appendix

\end{document}